\documentclass[a4paper,12pt]{article}

\usepackage{amsmath,amssymb,amsfonts}
\usepackage[dvips]{graphicx}
\usepackage{epsfig}

\usepackage{cite}
\usepackage{amsmath,amssymb,amsfonts,graphicx}
\usepackage{epsfig}
\usepackage[left=3cm,top=3.6cm,right=3cm,bottom=3.6cm,bindingoffset=0cm]{geometry}

\usepackage{pdfsync}

\newcommand{\be}{\begin{equation}}
\newcommand{\ee}{\end{equation}} 
\newcommand{\beq}{\begin{eqnarray}} 
\newcommand{\eeq}{\end{eqnarray}}

\newcommand{\p}{\partial}
\newcommand{\Tr}{{\rm Tr}}

\newcommand{\bea}{\begin{eqnarray}}
\newcommand{\eea}{\end{eqnarray}}
\def\Tr{ \hbox{\rm Tr}}

\def\de{\partial}

\def\half{\frac{1}{2}}

\numberwithin{equation}{section}

\begin{document}

\title{
\begin{flushright}\ \vskip -1.5cm {\small {IFUP-TH-2015}}\end{flushright}
\vskip 40pt
\bf{ \Large
NonAbelian Vortices,  Large Winding Limits \\ and Aharonov-Bohm Effects }
\vskip 20pt}
\author{S. Bolognesi$^{(1,2)}$, C. Chatterjee$^{(2,1)}$ and  K. Konishi$^{(1,2)}$\\[20pt]
{\em \normalsize
$^{(1)}$Department of Physics "E. Fermi", University of Pisa}\\[0pt]
{\em \normalsize
Largo Pontecorvo, 3, Ed. C, 56127 Pisa, Italy}\\[3pt]
{\em \normalsize
$^{(2)}$INFN, Sezione di Pisa,    
Largo Pontecorvo, 3, Ed. C, 56127 Pisa, Italy}\\[3pt]
{ \normalsize Emails: stefanobolo@gmail.com, chatterjee.chandrasekhar@pi.infn.it}\\
{ \normalsize  konishi@df.unipi.it}}
\vskip 10pt
\date{March 2015}
\maketitle

\begin{center}
{\large Abstract}
\end{center}

Remarkable simplification arises from considering vortex equations in the large winding limit.
This was recently used  \cite{Bolognesi:2014saa}  to display all sorts of vortex zeromodes, the orientational, translational, fermionic as well as semi-local,  and to relate them to the apparently distinct phenomena of  the Nielsen-Olesen-Ambjorn magnetic instabilities. 
Here we extend these analyses to more general types of BPS nonAbelian vortices, taking as a prototype a system with gauged $U_0(1)\times SU_{\ell}(N)\times SU_r(N)$ symmetry where the VEV of charged scalar fields in the bifundamental representation breaks the symmetry to $SU(N)_{\ell + r}$. The presence of the massless $SU(N)_{\ell + r}$  gauge fields in $4D$ bulk introduces  all sorts of non-local,  topological phenomena such as  the nonAbelian Aharonov-Bohm effects, which in the theory with global $SU_r(N)$ group  ($g_r=0$)  are  washed away by the strongly fluctuating orientational  zeromodes in the worldsheet.  Physics changes  qualitatively at the moment the right gauge coupling constant $g_r$ is turned on.

\newpage

\section{Introduction}

Vortex solutions at large winding limits \cite{wallvortex1,wallvortex2}  provide for an interesting theoretical laboratory, where properties characteristic of these soliton solutions can be exactly analyzed.  
For instance,  all types of zeromodes associated with nonAbelian vortices have recently been re-analyzed in detail and with great generality \cite{Bolognesi:2014saa}, which allowed to unearth the deep connection between two apparently unrelated physics phenomena of nonAbelian vortices \cite{Auzzi:2003fs,Hanany:2003hp,Shifman:2004dr} and  the Nielsen-Olesen-Ambjorn magnetic instabilities \cite{Nielsen:1978rm,Ambjorn:1988tm,Ambjorn:1989bd}, via universal mechanism of various field zeromodes in a critical magnetic field.

As is well known, vortex equations have not been analytically solved,  even in the simplest  case of the BPS saturated, or self-dual,  Abelian (Abrikosov-Nielsen-Olesen) vortex.\footnote{Analytic vortex solutions are however known in certain  systems defined on hyperbolic spaces with tuned curvature \cite{Manton:2009ja}-\cite{Eto:2012aa
}. For relation to nonAbelian vortices and large winding see for example \cite{Sutcliffe:2012pu}.} 
Therefore the study of these solutions in a somewhat idealized setting  of large-winding limits may be of some interest. 
Indeed, a remarkable feature is that the vortex equations reduce in this limit to certain algebraic equations. This simplification allows us  to exhibit vortex configurations explicitly in cases which have not been analyzed up to now,  thus enabling us to study  various aspects of vortex solutions so far little explored.

The purpose of this paper is actually   two-folds. The first is to extend our recent analysis  \cite{Bolognesi:2014saa} to more general systems, 
in particular, to the cases of fully  or partially gauged version of nonAbelian vortices.  A prototype example we choose \footnote{Extension to analogous models involving gauge groups such as $SO(N)$ or  $USp(2N)$ groups is quite straightforward. %\cite{...}. 
} 
is  $U_0(1)\times SU_{\ell}(N)\times SU_r(N)$ gauge theory, where the VEV of charged scalar fields in the bifundamental representation break the gauge symmetry to diagonal subgroup $SU(N)_{\ell + r}$.  We shall see that the simplification 
in the large winding limit allows us to determine the field configurations  concretely, and  consequently to find  the dimension of the vortex moduli space in all cases.  Our result suggests that the vortex moduli for the minimally winding solutions remains  $CP^{N-1} \times {\bf C}$, the latter being the translational modes, independent of $g_r$.   The analysis of the vortex configurations and the zeromodes is rather non-trivially extended to the case where only part of the $SU_r(N)$ symmetry is gauged.

The second purpose of this work is to  discuss what appears to be a  surprising  contrast and a sort of complementarity existing 
 between the physics of $g_r=0$  and  $g_r\ne 0$ systems, where $g_r$ is the $SU_r(N)$ coupling constant.  In the former case one has 
nonAbelian vortex with fluctuating, quantum $CP^{N-1}$ dynamics on the $2D$ string world sheet, extensively studied in the last ten years. Where as  in the latter,  the $CP^{N-1}$ orientational modes are immersed and interact with $4D$ degrees of freedom which remain massless.  
All sorts of  $4D$  non-local and  topology-related  phenomena arise,  such as   nonAbelian Aharonov-Bohm (AB) effects, nonAbelian Cheshire charge,  nonAbelian statistics, and so on, studied somewhat earlier in different examples \cite{Alford:1990ur,Alford:1990mk,Alford:1992yx,Lo:1993hp}. 
The discussion of the low-energy $2D-4D$  effective action in the general cases  of fully or partially gauged models will be presented  in a separate work. 

The paper is organized as follows. In Section \ref{uno} we introduce the theory of the fully gauged nonAbelian vortex. We then study in detail the large winding vortex of such theory, solve for the vortex profile  and compute the  zeromodes. In Section \ref{tre} we discuss the AB phenomena and non-local effects which arise for gauged vortices of this kind.  In Section \ref{quattro} we consider the case in which only a $U(1)$ part of the $SU_r(N)$ is gauged. A concluding discussion is in Section \ref{cinque}.

\section{Gauged nonAbelian vortices in the  BPS limit}
\label{uno}

In the $U_0(1)\times SU_{\ell}(N)\times SU_r(N)$ gauge theory we are interested in \footnote{The investigation of this system was initiated in \cite{Konishi:2012eq} in the approximation in which one of the nonAbelian gauge couplings, e.g., $SU_r(N)$, is arbitrarily weak.}, 
 two $SU(N)$ gauge groups, with gauge fields $A^{(\ell)}_{\mu}$ and $A^{(r)}_{\mu}$, act on  the left and on the right of the set of  $N \times N$ scalar fields $q$ as
\beq
q \to m_{\ell} \, q \, m_r^{\dagger} \ ,
\eeq
where $m_{\ell} \in SU_{\ell}(N)$ and $m_r \in SU_r(N)$.
Also,  an Abelian $U(1)$  factor   simply acts as  
\beq    q \to e^{i \alpha}  q\ ,
\eeq 
namely,  all scalar components have the unit charge.
The action in the BPS limit is the following
 \beq
{\cal L} &=& -\frac{1}{4}  \, f_{\mu\nu} f^{\mu\nu} -\frac{1}{2} \Tr\, (F^{(\ell)}_{\mu\nu}F^{(\ell)\mu\nu})  -\frac{1}{2} \Tr\, ( F^{(r)}_{\mu\nu}F^{(r)\mu\nu})  +  \Tr \, (D_{\mu} q)^{\dagger}(D^\mu q) \nonumber \\
 &&  -  \frac{g_0^2}{2}    \left[\Tr\, (q^{\dagger} q )   -  v_0^2\right]^2    -  \frac{g_{\ell}^2}  {2}  \sum_a    \left[ \Tr\, ( q^{\dagger}t^a q )\right]^2   - \frac{g_r^2}{2}  \sum_a   \left[ \Tr\, (q t^a  q^\dag )\right]^2\ , \label{BPSaction}
\eeq
with the covariant derivative given by
\beq
\label{covder}
D_{\mu} q = \partial_{\mu} q  - i  g_{0} a_{\mu} q - i  g_{\ell}  A^{(\ell)}_{\mu} q   + i  g_r q  A^{(r)}_{\mu}   \ . 
\eeq
The well studied case of  nonAbelian vortex with unbroken bulk $SU(N)$ flavor group  is recovered in the limit $g_r \to 0$ where the right gauge fields decouple (or equivalently $g_{\ell} \to 0$ where the left gauge fields decouple).
To simplify somewhat the formulas below we shall use below  also
\beq 
 \xi \equiv \frac{v_0^2}{N}  \ , \qquad \qquad e  \equiv  \sqrt{2N}   g_0\ ,
\eeq
so that for $g_r=0$, the limit of $U_{\ell}(N)$ theory correspond to $ e=g_{\ell}$.

The scalar field $q$ acquires a VEV in the vacuum.  By a gauge transformation, we can  bring to the form where it is proportional to the identity matrix 
\beq
\label{idq}
q=\left(\begin{array}{ccc}
\sqrt{{\xi}}&&\\
&\ddots&\\
&&\sqrt{{\xi}}\\
\end{array}\right) \ .
\label{qvacuum}
\eeq
An $SU(N)$ linear combination of the gauge fields remains massless
\beq
\label{masslesscombination}
{\cal A}_{\mu}=  A_{\mu}^{(unbroken)} =  \frac{1}{\sqrt{g_{r}^2 + g_{\ell}^2}  } \left( g_{r}   A^{(\ell)}_{\mu}   +     g_{\ell}  A^{(r)}_{\mu} \right)
\eeq 
 whereas  the orthogonal combination 
\beq
\label{massivecombination}
{\cal B}_{\mu} =  \frac{1}{\sqrt{g_{r}^2 + g_{\ell}^2}  } \left( g_{\ell}   A^{(\ell)}_{\mu}   -    g_{r}  A^{(r)}_{\mu} \right)    \label{brokenG}
\eeq 
and the $U(1)$ field $a_{\mu}$, are both  massive. 
%, with masses of the order of  $\sqrt{\xi}$.  
The main difference with respect to  the case where $SU_r(N)$ is a global symmetry,  is  the presence  of the  massless gauge bosons (\ref{masslesscombination})  propagating in the bulk.

The BPS completion is 
\begin{eqnarray}
T   & = &   \int d^{2}x \,  \left[ \,  \frac{1}{2} \left\{ f_{12}  + {g_{0}}  \left(\Tr\, (q^{\dagger} q)  -  N \xi\right)\right\}^{2} +   \Tr \left\{    \left( F^{(r)}_{12}  -  g_{r} \sum_a  \,t^a
 \Tr\, (q t^a  q^\dag)   \right)^{2} + \right.\right. \nonumber \\
  &&  +  \left. \left.   \left( F^{(\ell)}_{12} +  g_{\ell} \sum_a \, t^a \Tr\,  (q^{\dag} t^a  q)    \right)^{2}  \right\}   +    \Tr \,   |D_{1} q +   i D_{2} q|^{2} +   g_{0} N    \, \xi\,  f_{12} 
\right]  \ .
 \label{BPScompletion} 
\end{eqnarray}
The BPS equations are accordingly:
\begin{eqnarray}
   D_{1} q  +  i D_{2} q &=& 0 \ , \label{BPSequations1}\\
  f_{12}  +    {g_{0}}  \left(\Tr\, ( q^\dag   q) -  N \xi\right) &=& 0 \ ,   \label{BPSequations2} \\
      F^{(\ell) }_{12} + g_{\ell} \sum_a \, t^a  \Tr\, ( q^\dag    t^a   q )      &=& 0 \ ,  \label{BPSequations3} \\
        F^{(r)}_{12}  - g_{r}  \sum_a \,t^a  \Tr\, (q \,   t^a   q^\dag )     &=& 0 \ .   \label{BPSequations4}
\end{eqnarray}

\subsection{The vortex at large-winding order } 

Let us first consider a diagonal vortex with scalar profile
\beq  q =    \left(\begin{array}{cc}   q_1 e^{i n \varphi}  & 0 \\0 & q_2 {\bf 1}_{N-1}\end{array}\right)   \label{scalarbckgd}
\eeq
This, in the large winding limit, reduces to the simplified form
\beq  q &=&  v\, \left(\begin{array}{cc}0 &  \\ & {\bf 1}_{N-1}\end{array}\right)     \qquad   \qquad  r < R_{bag} \ ,\nonumber \\    q &=&    \sqrt{\xi} \left(\begin{array}{cc}e^{i n \varphi}   & 0 \\0 &  {\bf 1}_{N-1}\end{array}\right) \qquad   r>R_{bag} \ ,
\eeq
where $R_{bag}$ is the bag radius to be determined yet together with the scalar condensate inside the vortex bag $v$ which is in general different from $\sqrt{\xi}$.

For the moment only the fields that are needed for the vortex solution will be kept. 
They are  the Abelian gauge field and the  nonAbelian field in the broken part  ${\cal B}_{\mu}$ so that the covariant derivative is just
\beq   
D_{\mu} q = \partial_{\mu}q  -   i g_0 a_{\mu} q -  i g^{\prime}  {\cal B}_{\mu}^a  t^a  q 
\label{covsimpler}
\eeq
with the new coupling $g'$ defined by 
\beq
\label{gprimo}
g^{\prime} \equiv  \sqrt{g_r^2 + g_{\ell}^2} \ .
\eeq 
Also one can restrict to the Lie algebra component
\beq     t^{(N^2-1)} \equiv    \frac{1}{\sqrt{2N(N-1)} } \,  \left(\begin{array}{cc}N-1 &  \\ & - {\bf 1}_{N-1}\end{array}\right) \ .
\eeq

We now use the  the first BPS equation  (\ref{BPSequations1}). %$(\nabla_{1}+ i \nabla_{2} ) \,q =0$ 
Evaluated on the $N-1$ lower block proportional to the identity it  has the form
 \beq      g_0  a_i -    \frac{g^{\prime}}{\sqrt{2N(N-1)}}  {\cal B}_i^{(N^2-1)} =0 \ ,         \label{nowinding}
\eeq
where this is valid everywhere, both inside and outside the bag radius.
Evaluated on the first component, the same BPS equation gives 
\beq  g_0 a_i +   \frac{g^{\prime} (N-1) }{\sqrt{2N(N-1)}}  {\cal B}_i^{(N^2-1)} =  -   \epsilon_{ij} \frac{   r_j}{r^2} n A(r)  \label{winding}   \ ,
\eeq
where  the function $A(r)$ behaves as 
\beq      A(r)  =    \begin{cases}
   \frac{r^2}{R_{bag}^2}      &    \quad  r < R_{bag}    \\
    1  &  \quad   r>  R_{bag}\ .
\end{cases}      \label{Aofr}
\eeq
 Equations  (\ref{winding}) and (\ref{nowinding}) can be rewritten as equations for  the axial components 
   \beq      g_0  a_{\varphi} -   \frac{g^{\prime}}{\sqrt{2N(N-1)}}  {\cal B}_{\varphi}^{(N^2-1)} &=& 0
\ ,    \label{nowindingaxial} \\    g_0 a_{\varphi}+   \frac{g^{\prime} (N-1) }{\sqrt{2N(N-1)}}  {\cal B}_{\varphi}^{(N^2-1)}  &=&  \frac{n}{r}  A(r) \ ,    \label{windingaxial}
   \eeq
and thus the the solution is 
   \beq  
\label{phigauge} 
g_0 a_{\varphi} =    \frac{n}{N\, r}  A(r)\ , \qquad      g^{\prime} {\cal B}_{\varphi}^{(N^2-1)}  =   \sqrt{\frac{2(N-1)}{N}} \, \frac{n}{r}  A(r)\ ,  
\eeq
which, using the BPS equations (\ref{BPSequations3}) and (\ref{BPSequations4}), on the left and right components separately reads
\beq
    A^{(\ell, N^2-1)}_{\varphi}   &=&   \frac{g_{\ell}}{g^{\prime}}  {\cal B}_{\varphi}^{(N^2-1)}  =  \frac{g_{\ell}}{g^{\prime \, 2 }}   \sqrt{\frac{2(N-1)}{N}} \, \frac{n}{r}  A(r)\ ,  \label{lgauge} \nonumber \\
    A^{(r, N^2-1)}_{\varphi}   &=& -  \frac{g_{r}}{g^{\prime}}  {\cal B}_{\varphi}^{(N^2-1)}  =  -  \frac{g_{r}}{g^{\prime \, 2 }}   \sqrt{\frac{2(N-1)}{N}} \, \frac{n}{r}  A(r)\ .\label{rgauge}
\eeq
From (\ref{Aofr}) one sees that the gauge fields have the exact solenoid form
\beq     A^K_{\varphi}     
   =    \begin{cases}
      \frac{ c_K \, r}{R_{bag}^2}      &    \qquad  r < R_{bag}     \\
   \frac{ c_K}{r}    & \qquad   r>  R_{bag} \ ,
\end{cases}      \label{GaugeA}
\eeq
where the constant $c_K$ depends on the gauge field component, 
\beq    c_K= \quad  \frac{n}{g_0 N}\ , \qquad  \frac{   n  \, g_{\ell}}{g^{\prime \, 2 }}   \sqrt{\frac{2(N-1)}{N}} \ , \qquad  -    \frac{   n  \, g_{r}}{g^{\prime \, 2 }}   \sqrt{\frac{2(N-1)}{N}}\ , \quad 
\eeq
for  $K= U(1)$, $SU_{\ell}(N)^{N^2-1}$ and $SU_r(N)^{N^2-1}$, respectively.

Up to now only the first BPS equation (\ref{BPSequations1}) and the assumption of large winding are used.
On the other hand, the rest of the  BPS equations  (\ref{BPSequations2})-(\ref{BPSequations4})   give the magnetic fields directly. Inside the vortex bag ($r< R_{bag}$)  they are constant:
\beq    f_{12} &=&  % -   \frac{e}{\sqrt{2N}}   ((N-1) v^2 - N \xi) 
= -  g_0  ((N-1) v^2 - N \xi)  \ ,  \label{magfield0} \\
 F_{12}^{(r)} &=&
  - \frac{g_r  \sqrt{N-1} }{\sqrt{2N}} v^2  \,  t^{(N^2-1)}    \ ,  \label{magfieldr} \\
    F_{12}^{(\ell)}  &=& 
 \frac{g_{\ell}  \sqrt{N-1} }{\sqrt{2N}} v^2  \,  t^{(N^2-1)}    \ , \label{magfieldl}
\eeq
whereas they vanish identically  outside
\beq   f_{12}=0 \ , \qquad F_{12}^{(r)} =0 \ , \qquad  F_{12}^{(\ell)} =0,  \qquad r> R_{bag}\ . 
\eeq
 Note that these correspond to a  magnetic field for the `broken'  gauge fields ${\cal B}_{\mu}$  (\ref{massivecombination})
\beq   {\cal B}_{12} =       g^{\prime}  \frac{  \sqrt{N-1} }{\sqrt{2N}} v^2 \ ,\eeq
while the bulk massless field  ${\cal A}_{\mu}$  (\ref{masslesscombination})  does not carry any magnetic flux.

Before proceeding further  it is useful to remember that the gauge fields  (\ref{GaugeA})   and the corresponding magnetic fields are 
simply related as 
\beq     F_{12} =    \begin{cases}
      \frac{ 2   c  }{R_{bag}^2}      &    \quad  r < R_{bag}     \\
0   & \quad   r>  R_{bag}
\end{cases}    
\quad  \longleftrightarrow   \quad A_{\varphi}=
 \begin{cases}
      \frac{ c \, r}{R_{bag}^2}      &    \quad  r < R_{bag}     \\
   \frac{ c  }{r}    & \quad   r>  R_{bag}
\end{cases}  
    \label{GaugeA&M}\eeq
as can be easily checked.   The consistency  for the relative strengths of Abelian and nonAbelian  gauge fields as determined by (\ref{phigauge}) and
by  (\ref{magfield0})-(\ref{magfieldl})   then determine the scalar condensate inside the bag, $v$:\footnote{As a check,  Eq.~(\ref{expression})  reduces, in the standard $U_{\ell}(N)$ model (i.e.,  with $g_r=0$ and $g_{\ell} =e$),  to  $
v^2 = \xi,$     
whereas   in the case of $g_r=0$, $e \ne g_{\ell}$,  to   
\[    v^2 =   \frac{e^2 N \xi}{g_{\ell}^2 +e^2 (N-1)} \stackrel {N=2} {=}    \frac{ 2   e^2  \xi}{g_{\ell}^2 +e^2},
\]
which is precisely  Eq.(3.43) of  \cite{Bolognesi:2014saa}.}
\beq   
v^2=   \frac{g_0^2  N \xi}{(g_{\ell}^2+ g_r^2)/2N + g_0^2 (N-1)} =   \frac{e^2 \xi N}{  g^{\prime \, 2} + e^2 (N-1)}    \ .   \label{expression}
\eeq
For later use, we record the $U(1)$ magnetic field
\beq   
f_{12}= -  g_0  ((N-1) v^2 - N \xi) =\frac{e   N \xi  }{\sqrt{2N}}     \frac{g^{\prime \, 2}}{g^{\prime \, 2}  + e^2  (N-1)} \ .
\eeq

\subsection{Flux, bag radius and  vortex  tension  \label{flux}  }

Actually, there are even stronger conditions,  which determine the vortex radius  itself.   
This can be clearly seen if one recalls that 
the magnetic flux can be  either directly calculated from  $\varphi$ component of the gauge fields, (\ref{phigauge})-(\ref{rgauge}), or by integrating the constant magnetic fields inside the bag (\ref{magfield0})- (\ref{magfieldl}).  They must agree.

From  Stokes' theorem, one gets 
\beq   
\Phi=  \oint_{r> R_{bag}}      r   \, d \varphi  \, A_{\varphi}\ , 
\eeq
that is
\beq \label{fluxes} 
  \Phi^{(0)} =      \frac{ 2\pi n}{ g_0    N}  \ , \ \  &&
  \Phi^{(\ell)} =     \frac{g_{\ell}} {g^{\prime \, 2}}   \frac{\sqrt{2N(N-1)}}{N}   2\pi n\, t^{(N^2-1)} \ ,\nonumber \\ 
 &&  \Phi^{(r)} =  -    \frac{g_{r}} {g^{\prime \, 2}} \frac{\sqrt{2N(N-1)}}{N}   2\pi n \, t^{(N^2-1)}  \ . 
\eeq
On the other hand, by integrating the magnetic fields   (\ref{magfield0})-(\ref{magfieldl}) over the bag area, one gets 
\beq    \Phi^{(0)} =   \pi R_{bag}^2    f_{12}\ ,    \qquad \Phi^{(\ell)} =   \pi R_{bag}^2    F_{12}^{(\ell)}\ , \qquad  \Phi^{(r)} =   \pi R_{bag}^2    F_{12}^{(r)}\ .  \label{fluxesBis}
\eeq
Equating the results (\ref{fluxes}) and (\ref{fluxesBis})  one finds the vortex radius
\beq    R_{bag}^2 =      \frac{4 n}{g^{\prime \, 2}  v^2}\ .  \label{R1}
\eeq
Note that the  same result for $R_{bag}$ is found  regardless of which gauge field component ($U(1)$, $SU_{\ell}(N)$ or $SU_r(N)$)  is used to make the matching, as is expected. 
This is not really accidental:   the condition that the {\it relative strengths}   among  the various magnetic field tensors   (\ref{magfield0})-(\ref{magfieldl}) and   those among the 
gauge fields   (\ref{phigauge})-(\ref{rgauge})   be the same,   has been used to determine the value $v$ of the scalar condensate inside the vortex   (\ref{expression}).

The vortex tension is    ${g_{0} N }   \, \xi\, $  times the Abelian flux in our BPS system, (\ref{BPScompletion}), therefore is  equal to  
\beq  T= 2\pi n  \xi \ .   
\label{tension}
\eeq

In the large winding limit the BPS equations for the scalar and gauge fields are  thus all explicitly solved,  see (\ref{phigauge})-(\ref{rgauge}).  These results give directly the fluxes associated to the $U(1)$, $SU_{\ell}(N)$ and $SU_r(N)$ gauge fields  ((\ref{fluxes}).  On the other hand, the BPS equations give directly the non vanishing magnetic fields inside the vortex bag (they vanish outside), (\ref{magfield0})-(\ref{magfieldl}). The comparison between the flux and the magnetic field yields the bag radius. The result for the latter agrees with what one obtains from the balancing the vacuum potential energy and the magnetic field energy, to minimize the total energy inside the vortex,  as was  done in the original papers  on the large winding vortices \cite{wallvortex1,wallvortex2}, see also \cite{Bolognesi:2014saa}. 
The agreement is again not a mystery in our BPS system:  the equations of motion are the energy minimizing conditions.    
Thus the most powerful result is that in the large winding limit,  the BPS equations for the vortex configurations effectively reduce to simple algebraic equations. 
This simplification allows us to go further,  to analyze the field mixing inside the vortex, determination of the effective gauge field masses,  and the 
counting of the vortex zeromodes,  as will be done in the following.

\subsection{$W$ boson masses  and field mixing \label{Wmass}}

One of the difficulties in the analysis of gauged nonAbelian vortices  \cite{Konishi:2012eq} was the fact that inside the vortex the gauge field mixing varies with the distance from the vortex core.
Here this problem is dealt with  straightforwardly,   thanks to the simplification pointed out above.  
It is easier now to work in the singular gauge.
In the scalar background 
\beq q = 
\left(
\begin{array}{cc}
  q_1 & 0   \\
  0& q_2 {\bf 1}_{N-1} 
\end{array}
\right) \ ,  \eeq
the  scalar kinetic terms becomes 
\begin{eqnarray}
D_{\mu} q &=&   \p_{\mu} q - i  g_0 a_{\mu} q -i g_{\ell} A^{(\ell)}_{\mu} q + ig_r q A^{(r)}_{\mu}   \nonumber   \\
&=&   \p_{\mu} q - i g_0 a_{\mu} q -i g' {\cal B}^{N^2 -1}_{\mu} t^{N^2 -1} q - i g'   \sum_{a=1}^{(N-1)^2-1} {\cal B}^a_{\mu}  t^{a} q    \nonumber \\ 
&&   -\frac{i}{\sqrt{2}}  \left(g_{\ell} W^{\pm (\ell)}_{\mu} q - g_r q   W^{\pm (r)}_{\mu}\right)  \ , \label{scalarkin} 
\end{eqnarray}
where  \beq 
\label{corners} 
W^{\pm (\ell,r)}_{{\mu}}  = \frac{1}{\sqrt{2}}
\left(
\begin{array}{cc}
  0 &  W^{\dagger}_{\mu} \\
  W_{\mu}  &  0  
    \end{array}
\right)
 \label{W+-}
\eeq
and  $W_{\mu}$ is an $(N-1)$  component complex vector,
\begin{eqnarray}
W^{\dagger}_{\mu} = \left(W^{1 *}_{\mu} , W^{2 *}_{\mu}  , \cdots, W^{(N-1) *}_{\mu}\right)  
\end{eqnarray}
appearing in the  $(1,i)$, $(i,1)$   $(i=2,3, \ldots, N)$    corner of (\ref{corners}).  Eq.~(\ref{scalarkin})  shows that   
the $U_0(1)$ and $U(1)\times SU(N-1) \subset SU(N) $ part of the the broken $SU(N)$, ${\cal B}_{\mu}$ 
are all massive, with masses $g^{\prime} \sqrt{ \xi}$.    As for  the unbroken $SU(N)$,    the $U(1)\times SU(N-1) \subset SU(N) $ part of the multiplet $\mathcal{A}_{\mu}$   are massless, whereas  
 the   $(1,i)$, $(i,1)$ components (\ref{W+-})  mix with the same components of the broken gauge fields  ${\cal B}_{\mu}$  in a precise way, as follows.

\subsubsection{Mass eigenstates in the bulk}

 Outside the vortex  $q_1 = q_2 =\sqrt{\xi}$,  and  the square of the last term of  (\ref{scalarkin})  gives 
\begin{eqnarray}
     &&\frac{\xi}{2}  \sum^{N-1}_{i = 1}\left(| g_{\ell} W_{\mu}^{(\ell)} - g_r W_{\mu}^{(r)} |^2 + 
|g_{\ell} W_{\mu}^{(\ell)}  - g_r W_{\mu}^{(r)} |^2\right) \nonumber \\
&& = \xi \sum^{N-1}_{i = 1} {g'}^2 |{\cal B}^{-}_{\mu} |^2\ .
\end{eqnarray}
 So ${\cal B}_{\mu}$'s  are all massive (broken $SU(N)$)   with mass $g^{\prime}  \xi$,   whereas  $\mathcal{A}$'s are all massless (unbroken $SU(N)$).    The $U(1)$  gauge boson   $a_{\mu}$ is also massive.

\subsubsection{Mass eigenstates inside the vortex bag}  

Inside the vortex bag $q_1 =0$ , $q_2 = v \ne 0$, therefore $|D_{\mu} q|^2$ contains
 \begin{eqnarray}
= \frac{v^2}{2}  \sum^{N-1}_{i = 1}\left(| g_{\ell} W_{\mu}^{(\ell)} |^2 + 
|g_r W_{\mu}^{(r)} |^2\right) \ .
\end{eqnarray}
 So all the fields in the coset  $SU(N)/U(1)\times SU(N-1)$  (\ref{corners}) of the left and right $SU(N)$ are massive.  Note that 
  the left and right $W^{\pm}$ components are mixed differently outside and inside the vortex. Inside, the mass eigenstates are  $W^{\pm (\ell)}$ and $W^{\pm (r)}$, and this fact will turn out to be crucial for the determination of the vortex zeromodes below. 
  Outside the vortex (in the bulk), the $W^{\pm}$ components of ${\cal B}_{\mu}$'s (massive) and  $\mathcal{A}$ (massless) are the mass eigenstates. 
  The $U(1)\times SU(N-1)$ components of ${\cal A}_{\mu}$ remains massless everywhere,  inside and outside the vortex.

\subsection{Vortex zeromodes  \label{zeromodes}}

With this knowledge it is now possible to determine the vortex zeromodes. 
We  found above  that  inside the vortex bag,  the mass eigenstates are $W^{\pm (\ell)}$ and $W^{\pm (r)}$, with masses
\beq    m_{W^{(\ell)}}^2  =   \frac{g_{\ell}^2}{2}  v^2\ , \qquad     m_{W^{(r)}}^2  =   \frac{g_{r}^2}{2}  v^2  \ .\label{masses}
\eeq
As for the "magnetic fields"  felt by these $W$ bosons,  it is clear that $W^{\pm (\ell)}$ is coupled to  $F_{12}^{(\ell)}$ only, and $W^{\pm (r)}$  to  $F_{12}^{(r)}$ only, as they arise from the original Yang-Mills action.  
Working out the coupling of  $W^{\pm (\ell)}$, as in  \cite{Bolognesi:2014saa},  and similarly for   $W^{\pm (r)}$,  one finds the magnetic fields 
\beq    B^{(\ell)}=   -  \frac{g_{\ell}}{2N}   \left( (N-1) -  (-1)\right)  v^2 =   -      \frac{g_{\ell}}{2}   v^2\ ,   \qquad  B^{(r)}=     \frac{g_{r}} {2}    v^2\ .     \label{critical}
\eeq
These turn out to have precisely the critical values  for  the left and right $W$ bosons, respectively. 
%
%These explicit expressions in fact  allow us to determine the dimension of the vortex moduli space (the space of the vortex zeromodes) 
%exactly, following the analysis made in  \cite{Bolognesi:2014saa}.  In our model the relevant zeromodes are associated with those of the gauge fields and the scalar zeromodes. 
%Let us consider first  the zero-energy modes  of the lowest Landau level for  $W^{\pm (\ell)}$ and $W^{\pm (r)}$. 
  Note that for this calculation one must use the magnetic fields to which these fields are coupled,  (\ref{critical}), rather than   (\ref{magfieldr}) or (\ref{magfieldl}).  The degeneracy of the left $W^{\pm(\ell, r)}$  zeromodes is then   ($n$ is the  the winding number)
\beq      d_{\ell} =    (N-1) \frac{g_{\ell} B^{(\ell)}}{2}   \,   R_{bag}^2  =  (N-1)  \frac{g_{\ell}^2}{g^{\prime \, 2}}   \, n,    \label{leftWzeromodes}
\eeq  
and similarly for the right $W$:
\beq      d_{r} =  (N-1)   \frac{g_{r } B^{(r)}}{2}   \,   R_{bag}^2  = (N-1)   \frac{g_{r}^2}{g^{\prime \, 2}} \, n\ , \label{rightWzeromodes}
\eeq  
where $N-1$ is the number of the charged  $W^{\pm}$ boson components,  and the degeneracy of the lowest Landau level \cite{Bolognesi:2014saa} has been taken into account.
The  total  number of  the $W^{\pm (\ell,r)}$  boson  zeromodes is  then:  
\beq      d_{\ell} +    d_{r} =  (N-1) \, n\ .   \label{Wbosons}
\eeq

As for the scalar modes,  one finds, by generalizing the discussion of (3.49)  of \cite{Bolognesi:2014saa},  the tachyonic mass for the $(11)$ scalar,  
\beq   m_{(11)}^{2} =  -   g_{0}^{2}  \frac{g^{\prime \, 2} N \xi}{g^{\prime \,  2} + e^{2} (N-1) }=  -   \frac{g^{\prime \, 2} v^2}{2}    \ .  \label{tachyon}
\eeq
On the other hand, the $q_{11}$ scalar is coupled to the magnetic fields (\ref{magfield0})-(\ref{magfieldl}), with couplings  $g_{0}$, $g_{r}$ and $g_{\ell}$, respectively. 
Summing all the contributions one gets the effective magnetic fields (times respective coupling constants)  felt by  the $q_{11}$ field:
\beq   && \frac{e g_{0}}{\sqrt{2N}} \left((N-1) v^{2} - N \xi\right)  -   \left(  \frac{g_r^{ 2}(N-1)}{2N} +  \frac{g_{\ell}^{2}(N-1) }{2N} \right)\, v^{2} 
=  \nonumber \\ &&  \ \ =  \frac{g^{\prime \, 2} g_{0}^{2} \xi N  }
{g^{\prime \,  2} + e^{2} (N-1)}= \frac{g^{\prime \, 2} v^2}{2} \ ,  \label{couplings}
\eeq 
which is again precisely the critical strength of the magnetic field for the tachyonic scalar, $q_{11}$, with mass (\ref{tachyon}). This gives $n$ scalar zeromodes, after taking into account the lowest Landau level degeneracy \cite{Bolognesi:2014saa}. 
Summing to the $W$ boson zeromodes, (\ref{Wbosons}), one finds 
\beq      (N-1)\, n +n = N \, n  
\eeq
as the total number of the zeromodes.

This dimension of the vortex moduli space, $N n$,   is the same as the case with the global $SU_r(N)$ group (see Ref.\cite{Bolognesi:2014saa}  and references cited therein): 
it includes the translational modes, the deformation modes in which the $n$-winding vortex splits into vortices of lower winding vortices, and so on.  In particular,  in the limit of  $n$ far separated vortices of the minimum winding,  the dimension of the moduli, $N n= (N-1+1) \,n$  is consistent with the 
$CP^{N-1}$ internal orientational  moduli plus the translation for each. The conclusion of \cite{Eto:2012aa} agrees with this. 
The  overall $CP^{N-1}$ orientational modes  clearly arise from the breaking of the $SU(N)$  left-right diagonal symmetry,  unbroken in the bulk, but broken by the individual  vortex to $U(1)\times SU(N-1)$.

\subsection{Relation with the flavored vortex}

The covariant derivative (\ref{covder}) in terms of the new fields (\ref{masslesscombination}) %${\cal A}_{\mu}$ 
and (\ref{massivecombination}) %${\cal B}_{\mu}$ 
is 
\beq
D_{\mu} q = \partial_{\mu} q  - i  g_{0} a_{\mu} q - i  \frac{g_{\ell}^2}{\sqrt{g_{\ell}^2+g_{r}^2}}  {\cal B}_{\mu} q - i \frac{ g_{r}^2}{\sqrt{g_{\ell}^2+g_{r}^2}}  q  {\cal B}_{\mu} +\nonumber \\
\  - i  \frac{g_{\ell}g_{r} }{\sqrt{g_{\ell}^2+g_{r}^2}} \left( {\cal A}_{\mu} q  - q {\cal A}_{\mu} \right) \ .
\eeq
We see that $q$ transforms in the adjoint representation with respect to ${\cal A}_{\mu}$.
We can rewrite the gauge kinetic term in terms of the new combinations ${\cal A}_{\mu}$ and ${\cal B}_{\mu}$. At the quadratic level it splits exactly into two separate kinetic terms for ${\cal A}_{\mu}$ and ${\cal B}_{\mu}$ respectively but the non-linear terms contain interactions between the two fields; so there is no great simplification in rewriting in this way.

A simplification arises if we make the following two restrictions: first we consider only configurations in which ${\cal A}_{\mu}$ is set to zero and second we consider only configuration for which $q$ and ${\cal B}_{\mu}$ are diagonal. 
This is the case indeed for the vortex profile function we considered  before.
In this case the action is equivalent to the following one
\beq
{\cal L} &=& -\frac{1}{2} \Tr\,  ({\cal B}_{\mu\nu}{\cal B}^{\mu\nu}) -\frac{1}{4}  \, f_{\mu\nu} f^{\mu\nu}  + \Tr \, (D_\mu q)^{\dagger}(D^\mu q) \nonumber \\
&&-  \frac{g_0^2}{2}    \left[ \Tr\, (q^{\dagger} q)   -  v_0^2\right]^2    -  \frac{{g'}^2}{2}  \sum_a \left[ \Tr\, (q^{\dagger} t^a  q) \right]^2   \ ,
\eeq
with covariant derivative (\ref{covsimpler}) and couplings $g_0$ and $g^{\prime}$ as defined in (\ref{gprimo}).
This is nothing but the theory of the `ordinary' nonabelian vortex with $SU(N)$ gauge field ${\cal B}_{\mu}$ and $SU(N)$ flavor group. 
We stress that this reduction is valid only if the above conditions are satisfied, thus only if $q$, $q^\dag$ and ${\cal B}_{\mu}$ all commute. 
For example we can apply this shortcut to compute the vortex profile, or the wall separating the interior from the exterior phase of the large winding vortex \footnote{Interestingly the domain wall equations derived in \cite{Bolognesi:2014saa} have recently appeared in a different BPS system \cite{Canfora:2014jsa}.}, just using these effective couplings. 
We can not use this shortcut instead to compute the fluctuations around the vortex, and in particular the zeromodes.

\section{Aharonov-Bohm   \label{AB}} 
\label{tre}

A general theory has been developed in the past to describe vortices with  unbroken gauge group in the bulk and the related Aharonov-Bohm (AB) effects  \cite{Alford:1990ur,Alford:1990mk,Alford:1992yx,Lo:1993hp}.  
Our model fits into this generic framework, but it also provides a new example with some features that are not present in the specific examples considered earlier. 
This is mostly due to the presence of a residual symmetry which is both continuous and nonAbelian. 
We shall now briefly go through the general theory and, step by step,  see how this is realized in our particular model, putting a particular emphasis on the novel aspects appearing in our model.

In  general   one has a gauge group $G$  broken to a subgroup $H$ by the expectation value of scalar field $ q$  in some representation of  $G$.
Vortices are classified by the homotopy group $\pi_1 (G/H)$. This is a first coarse-grained classification, a  more refined structure will be discussed below. $G$ is taken to be the universal cover of the gauge group so that it is simply connected, i.e.  $\pi_1(G)=1$. 
In this way one can take into account any possible representation of the gauge group, besides the representation of $q$ which one needs to construct the vortex.
The topological classification is thus equivalent to counting the disconnected components of the unbroken group $H$, i.e.  $\pi_1 (G/H) = \pi_0(H)$.
In our example the universal cover of the gauge group is
\beq
G =  \widetilde{U_0(1)} \times SU(N)_{\ell} \times SU(N)_r
\eeq 
An element of $G$ is represented as a list of three elements
\beq
\label{conv}
(e^{i \alpha},\, m_{\ell}, \,  m_r)
\eeq
where $m_{l,r}$ are two $SU(N)$ matrices. 
$\widetilde{U(1)}_0$ is the universal covering of $U(1)_0$:   it can be identified with the real line parametrized by $\alpha$ with the $+$ operation. 
The scalar field  $q$ lives in the representation $(1, N , \bar{N})$ of the group $G$.
This means that the field $q$, written as a matrix, transforms under the action of (\ref{conv}) as follows:
\beq
\label{transf}
q \to e^{i \alpha} \, m_{\ell} \, q \,  m_r^{\dagger}
\eeq

The scalar field $q$ acquires a VEV  in the vacuum.  By a gauge transformation, one can  bring it to the form  proportional to the identity matrix (\ref{idq}).
The group $G$ is thus broken to a subgroup $H$ that leave $q$ invariant. Using (\ref{transf}) it is seen  that the condition for $q$ to be invariant is 
\beq
{\bf 1} = e^{i \alpha} m_{\ell}  m_r^{\dagger}
\eeq 
One way to parametrized the elements of $H$, using the convention of (\ref{conv}), is the following
\beq
\left(e^{\frac{2\pi i k}{N}}, m_{\ell}, m_{\ell} \,e^{\frac{2\pi i k}{N}} \right) \in H \subset G \qquad {\rm with}  \qquad  k \in Z
\eeq
Note that $m_r$ is equal to $m_{\ell}$ times an element in the center of the group and the same phase must appear in the $U(1)$ component.
This is the group
\beq
H =  SU(N) \times Z
\eeq
There are disconnected components in $H$ labeled by the integer $k$. 
Thus one infers that 
\beq
\pi_1 (G/H) = \pi_0(H) = Z
\eeq
Each disconnected component is an $SU(N)$ group, in particular for $k$ multiple of $N$ it is exactly the diagonal group $SU(N)_{l + r}$ defined by $m_r = m_{\ell}$.
One could have labeled the elements of $H$ in a different but equivalent way using $m_r$ 
\beq
\label{parar}
\left(e^{\frac{2\pi i k}{N}}, m_r \,e^{-\frac{2\pi i k}{N}} , m_r  \right) \in H \subset G \qquad {\rm with}  \qquad  k \in Z\;.
\eeq

Now consider the presence of a vortex. Far from the vortex one is in the vacuum and the gauge field does not have any curvature.  Nevertheless the vortex  can yield physical effects. A particle encircling  around the vortex, even if it  stays always at large distances from it, can aquire a non-trivial transformation due to the parallel transport.

From the computation done in the previous section, the gauge connection at large distances, if the vortex orientation is in that particular direction,  is given by (\ref{phigauge}),(\ref{lgauge}):
\beq
\label{potbisbis}
&& a_{\varphi} =    \frac{n}{g_0 N\, r}\nonumber \\  &&  A^{(\ell)}_{\varphi}   =  \frac{g_{\ell}n\sqrt{2(N-1)}}{g^{\prime \, 2 }\sqrt{N} r}   t^{(N^2-1)}   \nonumber \\  &&   A^{(r)}_{\varphi}   =  - \frac{g_{r}n\sqrt{2(N-1)}}{g^{\prime \, 2 }\sqrt{N} r}   t^{(N^2-1)}   
\eeq 
A parallel transport along a large loop around this vortex results in the following gauge transformation: 
\beq
\Gamma= \left(\Gamma_0,\Gamma_{\ell} , \Gamma_r\right) =  \left( \exp{\int ig_0 a_{\mu}, \, {\rm P} \exp{\int ig_{\ell} A^{(\ell)}_{\mu}}} , \, {\rm P} \exp{\int ig_r A^{(r)}_{\mu} } \right)
\eeq
which, for the specific potentials (\ref{potbisbis}),  reads
\beq
\label{firstgamma}
\Gamma=\left( e^{\frac{i2 \pi n}{N}}, 
\left(\begin{array}{cc} e^{ \frac{i2 \pi g_{\ell}^2 n (N-1)}{g^{\prime \, 2 }N}}  &  \\ &  {\bf 1}_{N-1} e^{- \frac{i2 \pi g_{\ell}^2 n }{g^{\prime \, 2 }N}} \end{array}\right), \left(\begin{array}{cc} e^{ -\frac{i2 \pi g_{r}^2 n (N-1)}{g^{\prime \, 2 }N}}&  \\ &  {\bf 1}_{N-1} e^{ \frac{i2 \pi g_{r}^2 n }{g^{\prime \, 2 }N}} \end{array}\right)  \right) \nonumber \\
\eeq
Since $q$ comes back to itself after parallel transport around a loop, one must check for consistency that $\Gamma$ is an element of $H$, i.e.,  that it leaves $q$ invariant. 
For this one just must check that $\Gamma_r^{\dagger} \Gamma_{\ell}$ belongs to the center of $SU(N)$.  Indeed, 
\beq
\Gamma_r^{\dagger} \Gamma_{\ell} = \left(\begin{array}{cc} e^{ \frac{i2 \pi  n (N-1)}{N}}&  \\ &  {\bf 1}_{N-1} e^{- \frac{i2 \pi n }{N}} \end{array}\right) = {\bf 1}_{N} e^{- \frac{i2 \pi n }{N}}
\eeq
where the relation (\ref{gprimo}) was used.
Thus it was shown that
\beq
\Gamma \in H \subset G
\eeq

\begin{figure}[h!]
\centerline{
\epsfxsize10cm \epsfbox{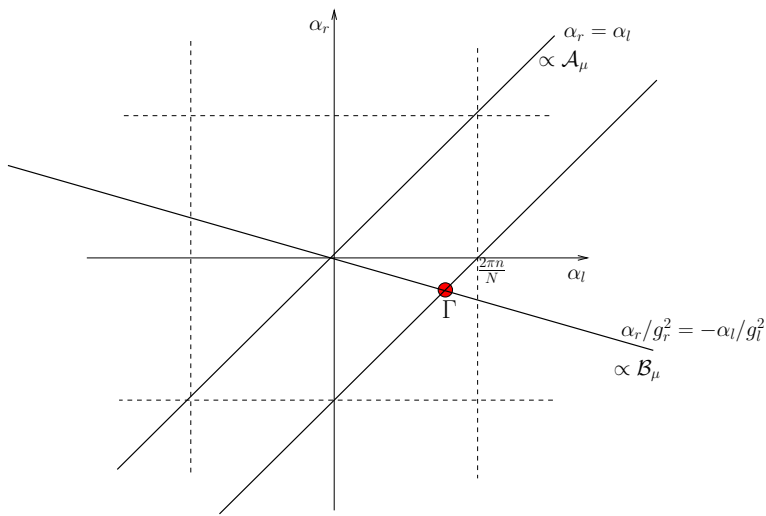}}
\caption{{\footnotesize A geometrical construction of the element $\Gamma$. }}
\label{torus}
\end{figure}

One can check this more explicitly using the geometrical construction of Figure \ref{torus}. 
Take a subgroup $U_{\ell}(1) \times U_r(1)$ defined by the elements 
\beq
(m_{\ell},m_r) = \left( \exp{i \alpha_{\ell} \sqrt{2N(N-1)}  \, t^{(N^2-1)}  }, \, \exp{i \alpha_{\ell} \sqrt{2N(N-1)}  \, t^{(N^2-1)}  } \right) 
\eeq
where $\alpha_{l,r}$ have period $2 \pi$.  In this basis the Cartan metric is proportional to the identity.\footnote{ This is not the  metric  inherited from the Yang-Mills terms which is proportional to the identity in the basis $(  A^{(\ell)}_{\mu},  A^{(r)}_{\mu})$ and where ${\cal A}_{\mu}$ and  ${\cal B}_{\mu}$ are orthogonal.} The direction of ${\cal A}_{\mu}$ corresponds to the line  $\alpha_{\ell} = \alpha_r$ while the direction of ${\cal B}_{\mu}$  corresponds to the line $ \alpha_{\ell}/g_{\ell}^2 = - \alpha_r/g_{r}^2 $. The element $\left(\Gamma_{\ell},\Gamma_r\right)$ lives in this plane and it is the intersection between the line  ${\cal B}_{\mu}$ and the line  ${\cal A}_{\mu}$ translated by $2\pi n /N$. 

To discuss the vortex orientation it is convenient to go into singular gauge. 
The parallel transport around the solution (\ref{potBis}) gives 
\beq
\label{gammatheta}
\gamma(\varphi) &=& \left(\gamma_0(\varphi),\gamma_{\ell}(\varphi) , \gamma_r(\varphi)\right) \nonumber \\
 & = & \left( e^{\frac{i \varphi n}{N}}, \left(\begin{array}{cc} e^{ \frac{i \varphi g_{\ell}^2 n (N-1)}{g^{\prime \, 2 }N}} &  \\ &  {\bf 1}_{N-1} e^{- \frac{i \varphi g_{\ell}^2 n }{g^{\prime \, 2 }N}}\end{array}\right), \left(\begin{array}{cc} e^{ -\frac{i \varphi g_{r}^2 n (N-1)}{g^{\prime \, 2 }N}} &  \\ &  {\bf 1}_{N-1} e^{ \frac{i \varphi  g_{r}^2 n }{g^{\prime \, 2 }N}}\end{array}\right)  \right) \nonumber \\
\eeq
Where $\gamma(2 \pi) = \Gamma$ of (\ref{firstgamma}). 
To go to the  singular gauge one fixes a direction, say $\varphi = 0$ where $q$ is in the identity form (\ref{idq}), and then perform a gauge transformation so that $q$ is everywhere the same. The required gauge transformation is the inverse of $\gamma(\varphi)$. This transformations cancel the gauge fields almost everywhere except along a singularity line (or surface) which may be taken to lie at $(\varphi = \pi; \, 0 \le  r < \infty$). The new fields outside the bag radius are
\beq
\label{potbis}
&& a_{\varphi} = 2\pi \delta(\varphi -\pi)   \frac{n}{g_0 N\, r}\nonumber \\  &&  A^{(\ell)}_{\varphi}   =  2\pi \delta(\varphi -\pi)  \frac{g_{\ell}n\sqrt{2(N-1)}}{g^{\prime \, 2 }\sqrt{N} r}   t^{(N^2-1)}   \nonumber \\  &&   A^{(r)}_{\varphi}   = -  2\pi \delta(\varphi -\pi)   \frac{g_{r}n\sqrt{2(N-1)}}{g^{\prime \, 2 }\sqrt{N} r}   t^{(N^2-1)}   \ ,
\eeq 
where $\delta(\varphi -\pi)$ is a Dirac delta function and $q$ is constant ((\ref{idq})) everywhere. If one computes the AB transformation in this new gauge one obtains the same result as in regular gauge (\ref{firstgamma}). The only difference is that here all the transformation is acquired when the loop crosses the singularity line (surface).  Clearly, while the AB phase is a physical observable  the position of the singularity surface ($\varphi=\pi$)  is a gauge choice, in perfect analogy with the position of the Dirac string emanating from a monopole. For completeness  the fields in the interior of the vortex bag in the singular gauge are give by:
\beq
\label{potBis}
&& a_{\varphi} =    \frac{n}{g_0 N \ r} \left(\frac{r^2}{R_{bag}^2}  -1 + 2\pi \delta(\varphi -\pi) \right) \nonumber \\  &&  A^{(\ell)}_{\varphi}   =  \frac{g_{\ell}n\sqrt{2(N-1)}}{g^{\prime \, 2 }\sqrt{N} r}   t^{(N^2-1)} \left(\frac{r^2}{R_{bag}^2}  -1 + 2\pi \delta(\varphi -\pi) \right)  \nonumber \\  &&   A^{(r)}_{\varphi}   =  - \frac{g_{r}n\sqrt{2(N-1)}}{g^{\prime \, 2 }\sqrt{N} r}   t^{(N^2-1)}   \left(\frac{r^2}{R_{bag}^2}  -1 + 2\pi \delta(\varphi -\pi) \right) \ .
\eeq

As the   exact $SU(N)$ gauge symmetry of the bulk  vacuum (\ref{idq})  is broken by an individual vortex to   $SU(N-1)\times U(1)$, there arise the global  $CP^{N-1}$  moduli of solutions.   The solution oriented in various directions in 
\beq  CP^{N-1}=   \frac{SU(N)}{ SU(N-1) \times U(1)}   \label{CPN-1}
\eeq
is  simply related to the solution before the rotation  (\ref{phigauge})-(\ref{rgauge})  by 
\beq    
&{\tilde q} =  U(B) \,q \,    U(B)^\dagger\ ,&    \nonumber \\ 
&{\tilde A}_i^{(\ell)} =  U(B) \,A_i^{(\ell)} \, U(B)^\dagger \ , 
\qquad   
{\tilde A}_i^{(r)}=  U(B) \, A_i^{(r)}  \, U(B)^\dagger \ , &\label{globalg}
\eeq 
where $B$ is a $N-1$ component complex vector parametrizing $CP^{N-1}$, Eq.~(\ref{CPN-1}), and the so-called reducing matrix $U(B)$    is given by
\beq   U(B) =  \left(\begin{array}{cc}1 & -B^{\dagger} \\0 & {\bf 1}_{N-1}\end{array}\right)  \, \left(\begin{array}{cc}x   & 0 \\0 & y^{-1}\end{array}\right)\left(\begin{array}{cc}1 & 0 \\B & {\bf 1}_{N-1}\end{array}\right) = \begin{pmatrix}
x^{-1} & - B^\dag y^{-1} \\
B x^{-1} & y^{-1}
\end{pmatrix} \ ,   \label{orientation}
\eeq
where $x$ and $y$ are a scalar and an $(N-1)\times (N-1)$ dimensional matrix, respectively:
\beq
x=  \sqrt{1 + B^\dag B }  \ , \qquad
y=  \sqrt{\mathbf{1}_{N-1} + B B^\dag  }\,\ .
\eeq
In the singular gauge this transformation does not change the value of $q$ outside the vortex but only inside. It does change instead the orientation of the gauge field along the singular line. 
For a generic vortex orientation one has 
\beq
\label{rotb}
\Gamma(B)&=&\left( e^{\frac{i2 \pi n}{N}}, U(B)\left(\begin{array}{cc} e^{ \frac{i2 \pi g_{\ell}^2 n (N-1)}{g^{\prime \, 2 }N}} &  \\ &  {\bf 1}_{N-1} e^{- \frac{i2 \pi g_{\ell}^2 n }{g^{\prime \, 2 }N}}\end{array}\right) U(B)^{\dagger}, \right. \nonumber \\
&& \qquad  \left. U(B) \left(\begin{array}{cc} e^{ -\frac{i2 \pi g_{r}^2 n (N-1)}{g^{\prime \, 2 }N}} &  \\ &  {\bf 1}_{N-1} e^{ \frac{i2 \pi g_{r}^2 n }{g^{\prime \, 2 }N}}\end{array}\right) U(B)^{\dagger}  \right)
\eeq
where $B$ and $U(B)$ are the same as defined before.
The element $\Gamma(B)$ is an object associated to each vortex and is a more refined quantity characterizing it  than the element of the homotopy group. For example $\Gamma(B)_0$ contains the information about the homotopy group but both $\Gamma_{\ell}(B)$ and $\Gamma_r(B)$ contain also the information about the orientation $B$. It will be seen below that this has physical, observable effects.

Now one  can  compute the AB effect of any particle around the vortex. 
First one has to specify the representation of $G$ of the particle, then find the element that correspond to $\Gamma(B)$ in this particular representation. 
For example a particle that transforms only as a fundamental of $SU_r(N)$, i.e. in the representation $(0,1,N)$ of $G$, is transformed by the matrix $ \Gamma_r(B)
$ in a parallel transport around the vortex. A  particle in the representation $(0,N,1)$ of $G$, is transformed by the matrix $ \Gamma_{\ell}(B)$.

The AB effect is always a unitary transformation. Every unitary matrix can be diagonalized in such a way that it  is a phase times the identity in every representation of the unbroken symmetry group. For example a particle  $(0,1,N)$ of $G$ corresponds just  to the unique representation $(N,0)$ of $H$ (using the parametrization (\ref{parar})). On the other hand the unitary matrix of the AB transformation is $ \Gamma_r(B)$  which is not proportional to the identity. This means that the actual symmetry group is not $H$ but a smaller group $\widetilde{H}$. $\widetilde{H}$ is defined as the elements of $H$ that commutes with $\Gamma$, that is 
\beq
\widetilde{H}  =  \frac{SU(N-1) \times U(1)}{Z_{N-1}} \times Z\;.
\eeq 
The representations of $\widetilde{H}$ are the ones which get an AB transformation which is proportional to the identity.  The fact that $H$ is broken to $\widetilde{H}$ is obvious when one  goes close to the vortex where there is magnetic flux. 
It is less trivial that this breaking can be detected even staying far from the vortex, where, at least  locally,  $H$ is unbroken.

Particles in the adjoint representation, such as $(0,{\rm Adj},1)$ or $(0,1,{\rm Adj})$, also get an AB transformation. These are particularly important because they are already present in the original theory:  they are the gauge bosons of the gauge group. Having a non trivial AB phase for gauge bosons may result in a topological obstruction, i.e. the impossibility of having a  continuously and globally extended basis for the Lie Algebra. 
The origin of the breaking  $H \to \widetilde{H}$, also observed at large distances, can be traced back to this fact.
The generators of $\widetilde{H}$ are the ones that remain unchanged in a parallel transport around the vortex.  The others may instead get a non-trivial phase when returned to the original point. 
For our model, fixing a particular vortex orientation as in (\ref{potBis}),  the parallel transport transformation is given by (\ref{gammatheta}).
Every element $h$ of the group $H$ is transformed by conjugation
\beq
h \to 
\gamma(\varphi)\, h \, \gamma(\varphi)^{-1}
\eeq
The elements in $\widetilde{H}$ remain invariant since they commute with 
$\gamma(\varphi)$. The discrete part $Z$ of $H$ belongs also to $\widetilde{H}$ so it does not have any non-trivial transformation. 
There are  $2(N-1)$ generators of the Lie Algebra $H$ which do not belong to $\widetilde{H}$. Due to the $SU(N-1)$ residual symmetry one  just has to compute one AB angle, which is the phase acquired, for example,  in a rotation among the following two generators
\beq
t^{1}  =   \frac{1}{2} \,  \left(\begin{array}{cccc} & 1& 0 &\dots  \\1 & && \\0 && &\\ \vdots &&& \end{array}\right) \;; 
\qquad 
t^{2} =  \frac{1}{2} \,  \left(\begin{array}{cccc} & -i& 0  & \dots \\i &  && \\0 &&&\\ \vdots &&& \end{array}\right)\;.
\eeq 
They are transformed as: 
%
%\beq 
%&&(0,t^{1}_{\ell},t^{1}_r) \longrightarrow (0,\gamma_{\ell}(\theta)\,t^{1}_{\ell} \, \gamma_{\ell}(\theta)^{-1} ,\gamma_r(\theta)\, t^{1}_r \,\gamma_r(\theta)^{-1} ) = \nonumber \\ 
%&&\ \ \ \ \ \ \ \ \ \  =    \left(0, \cos{\left(\frac{\theta g_{\ell}^2 n }{g^{\prime \, 2 }}\right)} t^{1}_{\ell}  - \sin{\left(\frac{\theta g_{\ell}^2 n }{g^{\prime \, 2 }}\right)} t^{2}_{\ell} , \cos{\left(\frac{\theta g_{r}^2 n }{g^{\prime \, 2 }}\right)}  t^{1}_r + \sin{\left(\frac{\theta g_{r}^2 n }{g^{\prime \, 2 }}\right)}  t^{r}_r \right)   \nonumber \\ \nonumber \\
%&&(0,t^{2}_{\ell},t^{2}_r) \longrightarrow (0,\gamma_{\ell}(\theta)\,t^{2}_{\ell} \, \gamma_{\ell}(\theta)^{-1} ,\gamma_r(\theta)\,t^{2}_r \,\gamma_r(\theta)^{-1} ) =
% \nonumber \\ 
%&& \ \ \ \ \ \ \ \ \ \  =    \left(0, \sin{\left(\frac{\theta g_{\ell}^2 n }{g^{\prime \, 2 }}\right)}  t^{1}_{\ell}  + \cos{\left(\frac{\theta g_{\ell}^2 n }{g^{\prime \, 2 }}\right)} t^{2}_{\ell} , - \sin{\left(\frac{\theta g_{r}^2 n }{g^{\prime \, 2 }}\right)}  t^{1}_r + \cos{\left(\frac{\theta g_{r}^2 n }{g^{\prime \, 2 }}\right)} t^{r}_r\right)  \nonumber \\
%\eeq
%
\beq 
&&(0,t^{1}_{\ell},t^{1}_r) \longrightarrow (0,\gamma_{\ell}(\varphi)\,t^{1}_{\ell} \, \gamma_{\ell}(\varphi)^{-1} ,\gamma_r(\varphi)\, t^{1}_r \,\gamma_r(\varphi)^{-1} ) = \nonumber \\ 
&&\ \ \ \ \ \  =    \left( 0,\,  \cos{\left(\tfrac{\varphi g_{\ell}^2 n }{g^{\prime \, 2 }} \right) } t^{1}_{\ell}  - \sin{\left(\tfrac{\varphi g_{\ell}^2 n }{g^{\prime \, 2 }}\right)}  t^{2}_{\ell},  \,
                                               \cos{\left(\tfrac{\varphi g_{r}^2 n }{g^{\prime \, 2 }} \right)}  t^{1}_r + \sin{\left( \tfrac{\varphi g_{r}^2 n }{g^{\prime \, 2 }} \right)}   t^{2}_r    \right)\;,   
                                                 \nonumber \\ \nonumber \\
&& (0,t^{2}_{\ell},t^{2}_r) \longrightarrow (0,\gamma_{\ell}(\varphi)\,t^{2}_{\ell} \, \gamma_{\ell}(\varphi)^{-1} ,\gamma_r(\varphi)\,t^{2}_r \,\gamma_r(\varphi)^{-1} )  =
 \nonumber \\ 
&& \ \ \ \ \ \  =    \left(0, \,  \sin{\left(\tfrac{\varphi g_{\ell}^2 n }{g^{\prime \, 2 }} \right)}  t^{1}_{\ell}  + \cos{\left(\tfrac{\varphi g_{\ell}^2 n }{g^{\prime \, 2 }} \right)} t^{2}_{\ell} , \, - \sin{\left(\tfrac{\varphi g_{r}^2 n }{g^{\prime \, 2 }} \right)}  t^{1}_r + \cos{\left(\tfrac{\varphi g_{r}^2 n }{g^{\prime \, 2 }} \right)} t^{2}_r  \right)  \;.
\eeq
Note that  the left and right generators  $t^{1,2}$  rotate with different angles. This is a reflection of the fact that the generators of $H$ which do not belong to $\widetilde{H}$ are changing, thus $H$ defined at different $\varphi$ is differrent.
At the end of a loop around the vortex one  comes back to the same $H$ but the left and right generators have rotated by the angles
\beq
\theta_{\ell} = \frac{2 \pi  g_{\ell}^2 n }{g^{\prime \, 2 }} \;,  \qquad \theta_r = - \frac{2 \pi  g_{r}^2 n }{g^{\prime \, 2 }}\;.
\eeq
One can verify  that one  returns to the original group $H$ since
\beq
\theta_{\ell} = \theta_r +  2 \pi n\;:
\eeq
the two angles are the same modulo $2 \pi$. 
But the angle $
\theta_{\ell}$, or equivalently $ \theta_r$, is in general non-trivial. This is the angle of topological obstruction.

Previous examples of nonAbelian strings with unbroken nonAbelian gauge group discussed in the literature  dealt mostly with some discrete $H$. One continuous and nonAbelian  example has been studied in \cite{Bucher:1993jj} where a GUT string carries a flux $\Gamma$ in an unbroken colored group $SU(3)$.  This has also been called `colored Alice string'. From the low-energy point of view it is quiet similar to our string. One difference is that the angle of topological obstruction discussed in \cite{Bucher:1993jj}   is $\pi$ whereas  in our case the angle can take any value between $0$ and $2 \pi$ depending on the coupling constants.

The group $\widetilde{H}$ contains the elements of $H$ which commute with $\Gamma$. This is the only part of $H$ that remains invariant under parallel transport around the vortex. 
The symmetry group is defined as the part of $H$ which can be globally defined, i.e. does not have any topological obstruction. In general the two definitions coincide, i.e. the symmetry group coincides with $\widetilde{H}$. Under very special circumstances the symmetry group may be larger. This happens if the angle of topological obstruction is multiple of $2 \pi$ but not necessarily zero. In our model this can happen if $g_{\ell}^2/g_{r}^2$ is a rational number.

The most interesting aspects of nonAbelian vortices are the long-range effects when  multiple vortices are present.  In the presence of only one vortex the AB effect is essentially Abelian, it just reduced to the problem of finding how the particle representation splits into representation of the unbroken group $\widetilde{H}$.
 With more vortices the  nonAbelian nature of these phenomena becomes  manifest.

%In the singular gauge where outside the vortex bag,   where
% the field configuration is that of the vacuum apart from a singular line which depart from the vortex center and goes to infinity.  In this gauge the AB phases of the particles going around a loop are all gained when crossing the singularity. 
  The  singular gauge is useful  when one considers the configuration with a collection of vortices of different orientations.  
   Due to the fact that the field reduces to the vacuum configuration outside the vortex,  there is no difficulty in patching together the configurations with more than one vortices.

 Let us consider the situation with two (parallel) vortices with different global orientations  $B_1$ and $B_2$.  If a particle encircles first  the $B_1$ vortex, and then   $B_2$ vortex, starting and ending at some reference point, $x_0$, it acquires an AB transformation given by 
 \beq  
  \Gamma_{ 1 \to 2} = \Gamma(B_2) \cdot   \Gamma(B_1),  \label{one} 
 \eeq 
evaluated in the particle representation. If it encircles the two vortices in the opposite order the phase will be
  \beq   \Gamma_{2 \to 1}  =     \Gamma(B_1) \cdot   \Gamma(B_2)\ . \label{two}  \eeq 
 As $ [ \Gamma(B_1),  \Gamma(B_2)] \ne 0$  
the  AB effect here is a nonAbelian type. This forms  a  representation of  the  (nonAbelian) first homotopy group, 
$\Pi_1(R^2/ \{x_1, x_2\}) $,   where $x_1$ and $x_2$ are the position of the two vortices.

Note that by a  global gauge transformation,  one of them, e.g., $\Gamma_{1\to 2}$, can be eliminated.  But then the contour $2 \to 1$ gets the phase,
\beq     \Gamma_{2 \to 1}^{\prime}  =     (\Gamma_{1 \to 2})^{-1}  \Gamma_{2 \to 1}  =         \Gamma(B_1)^{-1}  \cdot   \Gamma(B_2)^{-1}  \Gamma(B_1) \cdot   \Gamma(B_2)\ne 1\ .
\eeq
That is, the {\it ratio}  between the phases associated to the homotopic paths $1 \to 2$ and $2 \to 1$,   $  (\Gamma_{1 \to 2})^{-1} \cdot  \Gamma_{2\to 1} $,    is gauge invariant and is given by $  \Gamma(B_1)^{-1}  \cdot   \Gamma(B_2)^{-1}  \Gamma(B_1) \cdot   \Gamma(B_2)$, which is therefore an observable, physical effect.

\begin{figure}[h!]
\centerline{
\epsfxsize12cm \epsfbox{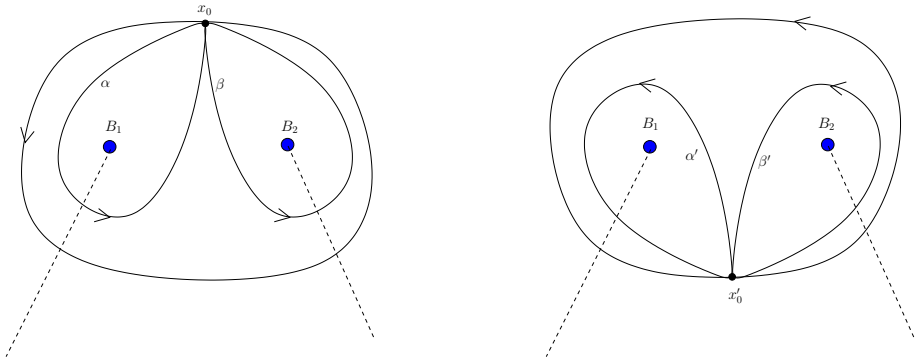}}
\caption{{\footnotesize Homotopy paths with two vortices. }}
\label{ud}
\end{figure}

Now the fact that  in our case the loop integral (\ref{firstgamma})  has a simple form, and the phase acquired does not obviously depend on the reference (the starting and ending) point $x_0$,     appears to bring us into paradoxical situations. One such paradox is as follows.  Namely if one encircles both of the vortices $B_1$ and $B_2$ starting from  a reference point above, $x_0$,   in the  anticlockwise direction, one expects that the particle gets the gauge transformation, 
$   \Gamma (B_2) \cdot \Gamma (B_1)
$,
as it is equivalent to the successive paths, $\alpha$ and $ \beta$  (see Fig. \ref{ud} left).     On the other hand, if the same closed contour is traced from a  starting  point  $x_0'$  below the vortices, in the same direction,   one appears to find another result, 
$    \Gamma (B_1) \cdot \Gamma (B_2)\ , 
$
(see Fig. \ref{ud} right), as the contour is now equivalent to the succession of the closed paths, $\beta^{'}$  first, and then $\alpha^{'}$.    This is a contradiction.  These  cannot be both  correct as the result should not depend on the precise position $x_0$,   and in particular  the notion such as `above' or `below' the vortices are not well defined.

Actually,   the association between a particular homotopic path and the relative AB phase must be made more carefully, as one is moving in a plane with branch cuts.   Suppose the reference point is  moved from $x_0$ to $x_0^{\prime}$, along a path which lies on the right of the vortex, $B_2$.  The path $\beta$ smoothly goes over to $\beta^{'}$,  whereas the path  $\alpha$ on the other hand is hooked by the vortex $B_2$ on the way: it is easy to see that  $\alpha$ ends up to become a path
$      (\beta^{'})^{-1}  \alpha^{'}   \beta^{'} $. 
   Let us fix the gauge so that  the AB phase associated with the path $\alpha$ is $\Gamma(B_1)$  
and that associated with $\beta$ is  $\Gamma(B_2)$;  the loop encircling the two vortices starting at $x_0$ then  gives an AB phase  $ \Gamma(B_2)  \Gamma(B_1) $.  If one starts at $x^{\prime}_0$, below the vortices,    while  the path  $\beta^{\prime}$  gives  the same phase as $\beta$ contour, $\Gamma(B_2)$,    the path  $\alpha^{'} =\beta \cdot \alpha \cdot \beta^{-1} $  does not give $\Gamma(B_1)$  but the conjugation
\beq        \Gamma_{\beta \cdot \alpha \cdot \beta^{-1} } =  \Gamma(B_2)  \Gamma(B_1) \Gamma(B_2)^{-1} \ .
\eeq
Therefore the full circle  yields this time 
  \beq      \Gamma(B_2)  \Gamma(B_1) \Gamma(B_2)^{-1}    \cdot  \Gamma(B_2)=   \Gamma(B_2)  \Gamma(B_1) \ ,
\eeq
the same  as the one  starting at $x_0$.  
Note that the two different AB phase  assignments for the loops  $\alpha$ and $\alpha^{\prime} $, both encircling the vortex $B_1$ once in the same direction,    are related to each other by a gauge transformation  depending on the vortex  $B_2$.

\section{Partial gauging: $U(1)_0 \times SU_{\ell} (N)\times U_r(1)$ theory }
\label{quattro}

Let us  now  come back to the problem of extending the first part of this paper, the detailed and concrete analysis of 
the vortex configurations and the determination of the zeromodes,  to the case the right $SU_r(N)$ symmetry is only partially gauged.
Let us now for concreteness  the case  in which only a $U(1)$ subgroup of $SU_r(N)$  is gauged. 
This  case is interesting because the right gauge interactions break explicitly the  $SU(N)_{\ell+r}$ symmetry of the bulk.
The $N=2$ ($SU_{\ell}(2)$)  case in which the right $U_r(1)$ gauge interactions are arbitrarily weak was studied in \cite{Evslin:2013wka};  here we 
study the same system but with an arbitrary right gauge coupling $g_r$ and in the large winding limit, and for  generic $SU_{\ell}(N)$.    
The starting point is the action 
 \beq
&&  {\cal L} =-\frac{1}{4}  \, f_{\mu\nu} f^{\mu\nu}  -\frac{1}{2} \Tr \, (F^{(\ell)}_{\mu\nu}F^{(\ell)\mu\nu})  -\frac{1}{2} \Tr \,( F^{(r)}_{\mu\nu}F^{(r)\mu\nu})  +  \Tr \,  (D_{\mu} q)^{\dagger}(D^\mu q) \nonumber \\
 &&  -  \frac{g_0^2}{2}    \left[\Tr\, (q^\dag  q)    -  N \xi \right]^2    -  \frac{g_{\ell}^2}  {2}   \sum_a  \left[ \Tr\,(q^{\dagger} t^a  q)\right]^2   - \frac{g_r^2}{2}   \left[ \Tr\, (q  t^{(N^2-1)}  q^\dag )\right]^2   \label{BPSactionpPG}
\eeq
with covariant derivative
\beq
D_{\mu} q = \partial_{\mu} q - i  g_{0} a_{\mu} q  - i  g_{\ell}  A^{(\ell)}_{\mu} q   + i  g_r q  A^{(r)}_{\mu}    \label{covariant}
\eeq
 and the BPS equations, but now the right gauge field has only one (color) component which we  take it in the direction
\beq    A^{(r)}_\mu =      t^{(N^2-1)}  \,  R_\mu\ .
\eeq
The scalar VEV has the same diagonal form as before (\ref{idq}), 
and an Abelian  $U(1)$ linear combination of the gauge fields remains massless in the bulk
\beq
\label{masslesscombinationBis}
{\cal A}_{\mu}=  A_{\mu}^{(unbroken)} =  \frac{1}{\sqrt{g_{r}^2 + g_{\ell}^2}  } \left( g_{r}   A^{(l,N^2-1)}_{  \mu}   +     g_{\ell}  A^{(r)}_\mu \right) \ , 
\eeq 
whereas other combinations the  $U(1)$ field $a_{\mu}$,  $A^{(l,a)}_{ \mu}$ for $ a\ne N^2-1$ and 
\beq
\label{massivecombinationBis}
{\cal B}_{\mu} =  \frac{1}{\sqrt{g_{r}^2 + g_{\ell}^2}  } \left( g_{\ell}   A^{(l, N^2-1)}_{\mu}    -    g_{r}  A^{(r)}_{\mu} \right) 
\eeq 
are all  massive.

The BPS completion can still be written down:
\begin{eqnarray}
T   & = &   \int d^{2}x\,  \Big[ \,  \frac{1}{2} \left\{ f_{12}  + {g_{0}}  \left(\Tr\, ( q^\dag   q) -  N \xi\right)\right\}^{2}       \nonumber \\ 
  & + &    \Tr \left\{ \left( F^r_{12}  -  g_{r}  \,t^{(N^2-1)}  \Tr\,( q  t^{(N^2-1)} q^\dag)  \right)^{2} +  
   \left( F^{(\ell) }_{12} +  g_{\ell} \sum_a \, t^a  \Tr\,(q^\dag  \,  t^a   q )  \right)^{2}  \right\}      \nonumber  \\    
  & + &  | D_{1} q +  i D_{2} q|^{2} +   {g_{0} N }   \, \xi\,  f_{12}    \Big]  
%  -  \left.\frac12\epsilon_{ij}\partial_{i}\left(i \nabla_{j}Q\bar Q- Q\,i\nabla_{j}\bar Q \right) \right\}
 \label{BPScompletionBis} 
\end{eqnarray}
and the BPS equations are the same as (\ref{BPSequations1}), (\ref{BPSequations2}), (\ref{BPSequations3}), while (\ref{BPSequations4}) is replaced by:
\beq
  F^{(r)}_{12}  - g_{r}  \,t^{(N^2-1)}  \Tr\,( q  t^{(N^2-1)} q^\dag)  = 0 \ .  \label{BPSequations44}
\end{eqnarray}

\subsection{Diagonal solution  \label{diagonalsol}} 

For the diagonal vortex oriented in the direction parallel to the $t^{(N^2-1)}$  right  and left gauge fields one has  the scalar field in diagonal form as in (\ref{scalarbckgd}),
and consequently  the solution is {\it identical} to those found in Section \ref{uno},  as the non diagonal components of $A^{(r)}_i$ (which do not exist!)  do not contribute. 
In particular the magnetic fields are given by (\ref{magfield0})-(\ref{magfieldl}), the VEV $v$ is (\ref{expression}), and the bag radius is (\ref{R1}).

\subsubsection{Gauge and scalar zeromodes around the diagonal solution  \label{zeromodesBis}}

In order to find the dimension of the moduli space
%, it is sufficient to consider the dimension of the zeromodes around any given point of the moduli space. 
%For simplicity 
we take the diagonal solution, and study the gauge-field and scalar-field  fluctuations around it, \`a la Olesen-Ambjorn, as done in \cite{Bolognesi:2014saa}. 

To start with, the background vortex configuration is identical to the diagonal solution (\ref{magfield0})-(\ref{magfieldl}). There are however some differences. 
The first  difference as compared to the discussion in Subsection~\ref{zeromodes}  is that now there are no nondiagonal gauge fields $ A^{(r)}_{\mu}$:    no $W^{ \pm (r)}$ gauge bosons
and hence   no zeromodes associated to them.  On the other hand, the left $W$ fields are coupled to the same background magnetic fields, thus the number of the left gauge zeromode are given precisely by (\ref{leftWzeromodes}).   Also, the scalar zeromodes associated with the $(\delta Q)_{11}$ field is again precisely given by $n$.

There is however another important difference as compared to the fully gauged case.    Expanding  the scalar potential
\beq   
V_{scalar}=   \frac{g_0^2}{2}    \left[\Tr\,( q^\dag  q)    -  N \xi \right]^2    +   \frac{g_{\ell}^2}  {2}   \sum_a  \left[ \Tr\,(q^{\dagger} t^a  q) \right]^2  +   \frac{g_r^2}{2}    \left[ \Tr \,(q  t^{(N^2-1)}  q^\dag) \right]^2   \label{potrgt}
\eeq
around the background
  \beq q = q^{diag} +
  \delta q,  \qquad    q^{diag} = 
\left(
\begin{array}{cc}
  0   & 0   \\
  0& v{\bf 1}_{N-1}
\end{array}  \label{expand}
\right) \ ,
\eeq
where $v$ is the same as  (\ref{expression}), 
%\beq   v^2=   \frac{g_0^2  N \xi}{(g_{\ell}^2+ g_r^2)/2N + g_0^2 (N-1)} =   \fr%ac{e^2 \xi N}{  g^{\prime \, 2} + e^2 (N-1)}\ ,   \label{expressionBis}
%\eeq
one finds that there are other tachyonic scalar field components. 
In the fully gauged model, these modes were `eaten' by the right and left gauge fields and their
associated zeromodes were properly taken into account as  $W^{(r)}$ zeromodes.  Let us write 
\beq  \delta q= \left(\begin{array}{cc}q_{11} & q_{1 i} \\q_{i1} & q_{ij}\end{array}\right)\ .
\eeq
By inserting (\ref{expand}) into (\ref{potrgt}),  one finds the negative mass square  terms,  
\beq  - \frac{ g^{\prime \,  2} v^2 }{2} \, |q_{11}|^{2} -    \frac{ g_r^{2}\, v^2 }{2} \, \sum_{i=2}^{N}  |q_{i 1}|^{2} \ ;
\eeq
all other terms are nonnegative.

On the other hand,  the magnetic fields to which  $q_{11}$ and $q_{i1}$ fields are coupled can be read off from (\ref{magfield0})-(\ref{magfieldl}).   They turn out to be
\beq     && B_{11} =\frac{ e^{2} g^{\prime \, 2} \xi  N}{2 \left(g^{\prime \, 2} + e^{2} (N-1)\right)} =    \frac{ g^{\prime \,  2} v^2 }{2} \ , \nonumber  \\ 
&& B_{i1}= \frac{ e^{2} g_{r}^{ 2} \xi  N}{2 \left(g^{\prime \, 2} + e^{2} (N-1)\right)}= \frac{ g_r^{2}\, v^2 }{2} \ , \qquad i=2,3,\ldots, N
\eeq
(these include the coupling constants, as in Eq.~(\ref{couplings})) which are in fact the critical values respectively for the tachyonic  $q_{11}$ and  $q_{i1}$ fields.  They represent thus exact zero-energy  modes, after taking into account the lowest Landau-level energy \cite{Bolognesi:2014saa}. The $(N-1) \,n $ zeromodes associated with the $q_{i1}$ fields thus  exactly compensate the missing  $W_r^{\pm}$ boson zeromodes,  compared  to the fully gauged case studied earlier.

Summarizing,  the total number of the zeromodes is  (see  Eqs.~(\ref{leftWzeromodes}),  (\ref{rightWzeromodes}))
\beq     d_{\ell} +   d_{r} + n =  N\, n\;, 
\eeq
where now the second and the third terms refer to the scalar modes:  the total is the same as in the fully gauged model of Section 2.

In conclusion one finds, by going to the limits of $n$ far separated minimum winding vortices,  $N$  zeromodes per each  of them.
This is consistent with a $CP^{N-1}$  moduli space, plus a translational mode.  The conclusions of \cite{Evslin:2013wka}  agree with this.

\subsection{Orthogonal solutions}

Another type of solutions exist, with   
\beq   
\label{assumption}
A^{(r)}_\mu  \equiv  0\;,
\eeq
and  with the same   $a_\mu$ and $A^{(\ell)}_\mu$  configurations as in the theory with global $SU_r(N)$ symmetry  (i.e., theory with $g_r=0$).   The scalar fields inside the vortex bag  take the form, 
\beq    q =   U  \, q^{diag}  U^\dagger  \label{rotatedsol}, \qquad   q^{diag}  =  \left(\begin{array}{cc}0 & 0 \\0 & v'  {\bf 1}_{N-1}\end{array}\right)\, . \label{constant}
\eeq
 More explicitly
\beq   q =   U  \, q^{diag}  U^\dagger  =  v^{\prime} \,  \left(\begin{array}{cc}B^\dagger B X^{-1}   & -B^\dagger  Y^{-1} \\-B X^{-1} & Y^{-1}\end{array}\right)\ .
\eeq
Both $v^{\prime}$ and a restriction on the possible orientation $B$ are determined below from the vortex equations.
The appropriate matrix of rotation $U$  (see Eq.~(\ref{orientation}))   can be found from the fourth BPS equation (\ref{BPSequations44})
and by requiring the right magnetic field to be zero (\ref{assumption}):
\beq     
0= R_{12} = g_r\, \Tr \, \left( q^{diag} U^\dagger    t^{(N^2-1)}   U\,{q}^{diag \dag }\,\right) \ .
\eeq
From 
\beq   q^{diag \dag} \,{q}^{diag} =  \left(\begin{array}{cc}0 &  \\ & v^{\prime \, 2} \, {\bf 1}_{N-1} \end{array}\right)\ ,
\qquad   
\eeq
one obtains
\beq  0=    R_{12} = 
   \frac{g_r   v^{\prime \,  2}}{\sqrt{2N(N-1)}}      
 \frac{ |B|^2 -   (N-1) }{1 + |B|^2} \ ,     \label{right}
   \eeq
and this  means that $B$ has a fixed modulus
   \beq       |B|^2=N-1\ .    
   \eeq
$F^{(\ell) }_{12}$  can be found from the    
the third BPS equation (\ref{BPSequations3}) 
\beq   F^{(\ell) }_{12} = -     g_{\ell} \, t^a  \Tr\,\left(  {q}^{diag \dag}  U^\dagger  \,  t^a  U   q^{diag}  \right)\ ,     \label{actually}
\eeq
which gives
\beq     (F^{(\ell) }_{12} )_{ij} =  v^{\prime \,2}  \left[ -  \frac{N-1}{2N} \delta_{ij} + \frac{1}{2}    \left(\begin{array}{cc}B^\dagger  y^{-2} B  & - B^\dagger y^{-2}  \\-y^{-2}  B  & y^{-2}    \end{array}\right)^T_{ij} \, \right]\ , \label{leftsln}
\eeq
where use was made of 
\beq  
 U   q^{diag}   {q}^{diag \dag}  U^\dagger =  v^{\prime \,2} \,  \left(\begin{array}{cc}B^\dagger  Y^{-2} B  & - B^\dagger Y^{-2}  \\-Y^{-2}  B  & Y^{-2}    \end{array}\right)\ , \eeq 
and of the identity
\beq 
( t^a )_{ij}    ( t^a )_{kl}  = - \frac{1}{2N}  \delta_{ij} \delta_{kl} + \frac{1}{2}      \delta_{il} \delta_{jk}\ .
\eeq
Actually,  (\ref{leftsln}) is  the same as the  diagonal  solution (\ref{magfieldl}) rotated by $U$ 
except for a change in the value of $v$:  
\beq      (F^{(\ell) }_{12} )_{ij}  =  U\, (F^{(\ell)\, diag }_{12} )_{ij} |_{v \to v'} \, U^\dagger\ .
\eeq 
%the left gauge fields $A^{(\ell)}_\mu$ is just the $SU_{\ell}(N)$ gauge transformed form of $A^{(\ell)}_\mu$  in the diagonal solution.  
Now it is easy to see how the first BPS equation is satisfied. Inside the vortex bag,  where the scalar fields take a constant form, (\ref{constant}), $(\de_1+ i \de_2 ) q=0$ and so the remaining terms of the equations become
\beq   
0 &=& \left( g_0 a_\varphi +  g_{\ell}\, A^{(\ell)}_\varphi   \right)\,  q  \nonumber \\
 & =&   U\,  \left( g_0 a_\varphi +  g_{\ell}\, A^{(\ell)}_\varphi   \right)^{diag} \, U^\dagger \,  U  \, q^{diag} \, U^\dagger \nonumber \\
& =&  U\,  \left( g_0 a_\varphi +  g_{\ell}\, A^{(\ell)}_\varphi   \right)^{diag} \,  q^{diag} \, U^\dagger\ ,   
\eeq
where 
  \beq   a_{\varphi}^{diag} =    \frac{n}{N\, g_0\, r}  A(r),\qquad     A^{(\ell)\, diag}_{\varphi}   =    \frac{1}{N\, g_{\ell}}   \, \frac{n}{r}  A(r) \, \left(\begin{array}{cc}N-1 &  \\ & -{\bf 1}_{N-1}\end{array}\right)\   .  \label{consistency}
\eeq    
This is solved by  
\beq         A(r)  =    \begin{cases}
   \frac{r^2}{R_{bag}^{'\, 2}},        &  \quad     r < R_{bag}^{'}  \;;   \\
    1,  & \quad   r>  R_{bag}^{'} \;.
\end{cases}     \eeq
so that 
\beq   &&  \left( g_0 a_\varphi +  g_{\ell}\, A^{(\ell)}_\varphi   \right)^{diag}  =    \left(\begin{array}{cc}  \frac{n}{r}  A(r) &  \\ & {\bf 0}_{N-1}\end{array}\right) \;,  \nonumber \\
 &&  .^.. \qquad      \left( g_0 a_\varphi +  g_{\ell}\, A^{(\ell)}_\varphi \right)^{diag}  q^{diag}= 0\,.   
\eeq
Consistency between (\ref{consistency})   with the remaining BPS equations (\ref{BPSequations2}) and (\ref{BPSequations3}):
\beq    f_{12} =   -   \frac{e}{\sqrt{2N}}   \left((N-1) v^{\prime \, 2} - N \xi\right) \;, %=   -  g_0  ((N-1) v^{' \, 2}   - N \xi)  ;  \label{magfield000} 
\nonumber \\
%\eeq
%\beq    
F_{12}^{(\ell)} = %    \frac{g_{\ell}}{2N} v^{'\, 2}      \left(\begin{array}{cc}N-1 & 0 \\0 & -{\bf 1}_{N-1}\end{array}\right) 
 \frac{g_{\ell}  \sqrt{N-1} }{\sqrt{2N}}v^{\prime \, 2}     \,  t^{(N^2-1)}    \ ,  \label{magfieldll}
\eeq
give   the last   relations
\beq   R^{\prime\, 2} =  \frac{4n}{g_{\ell}^2 \, v^{\prime \, 2}}\ ,   
\qquad   \qquad      v^{\prime \, 2} =    \frac{  e^2 N \xi} {g_{\ell}^2 +   e^2 (N-1)} \;, \label{Rorto}
\eeq
that complete the solution. 
Note that both the scalar condensate  inside the bag and the vortex bag radius itself    of these solutions are different from those of the diagonal solution, (\ref{expression}) and (\ref{R1}).
The $U(1)$ magnetic field is, in this case
\beq   
f_{12}= 
%-  g_0  ((N-1) v^{\prime \, 2} - N \xi) =  - g_0  N \xi    \left[   \frac{  e^2 (N-1) } {g_{\ell}^2 +   e^2 (N-1)}      -  1 \right]   \nonumber  \\ 
%&&  =g_0  N \xi    \left[  \frac{ g_{\ell}^2  } {g_{\ell}^2 +   e^2 (N-1)}       \right]=   
\frac{e   N \xi  }{\sqrt{2N}}    \left[  \frac{ g_{\ell}^2  } {g_{\ell}^2 +   e^2 (N-1)}       \right] \ .
\eeq

\subsubsection{Zeromodes around an orthogonal solution}  
The zeromodes around orthogonal vortices are similar to $g_r = 0$ case.  The degeneracy of the  $W^{\pm({\ell})}$  zeromodes is then  
% ($n =$  the winding number)
\beq      
d_{\ell} =     (N-1)   \, n,    \label{leftWzeromodespartial}
\eeq  
where $N-1$ is the number of the charged $W^{\pm({\ell})}$  bosons,  and the degeneracy of the lowest Landau level \cite{Bolognesi:2014saa} has been taken into account.

As for the scalar modes,  one finds a tachyonic mass for the  scalar,  
\beq   m^2_S =  -\half g^2_\ell v^{\prime \, 2}
%-\frac{e^2 g^2 \xi}{e^2 + g^2}\;,
\eeq
which becomes massless after the lowest Landau energy is taken into account. 
Summing to the $W^{\ell}$ boson zeromodes  (\ref{leftWzeromodespartial}), taking into account the Landau level degeneracy, one finds
\beq      (N-1)\, n +n = N \, n   \label{dimension}
\eeq
as the total number of the zeromodes.

\subsection{Absence of solutions interpolating $B=0$ and $|B|^2=N-1$ vortices: a puzzle? }

The vortex equations are simple algebraic equations at the large winding limit, as we have emphasized several times already.  It is in fact quite easy to convince oneself that 
there are no solutions of  the BPS equations, continuously interpolating between the $B=0$ and $|B|^2= N-1$ solutions.  This can be verified explicitly numerically in the case of $N=2$.  It is clear in general  that such a set of  solutions cannot exist  as the bag radius of the $B=0$ solution and that of the orthogonal ($|B|^2=N-1$) solution, (\ref{Rorto}),  are different.  

This brings us to a paradoxical situation. On the one hand, 
the dimension of the  vortex moduli space,  $N n$, is certainly suggestive of a $CP^{N-1}$ orientational moduli space of vortices for a singly winding vortex  ($N-1$ for internal moduli
and $1$ translational moduli), 
and indeed in the case of minimum vortices in $U_0(1) \times SU(2) \times U_r(1)$ theory, such continuous set of degenerate solutions have been explicitly constructed  
\cite{Evslin:2013wka}.   On the other hand, in the case of large-winding vortices, there seem to be no continuous set of solutions interpolating from $B=0$ to $|B|^2=N-1$ solutions. 
What went wrong?

\begin{figure}
\begin{center}
\includegraphics[width=5in]{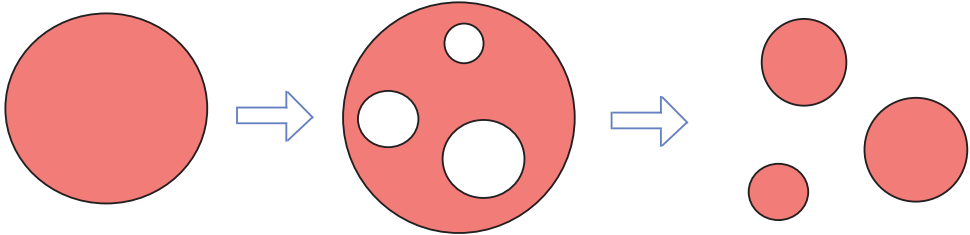}
\caption{A $n$ winding vortex all in the same orientation  (the left figure) is degenerate with the hole-vortex mixture  (the center) and the  far-separate vortices with smaller winding numbers $n_1+ n_2+\ldots =n$  (the right figure).}
\label{holes}
\end{center}
\end{figure}

A hint for a possible resolution of the puzzle comes from the idea of the ``hole vortex"  discussed in \cite{Bolognesi:2014saa}.  In the case of  $g_r=0$  theory (hence the case of 
nonAbelian vortex with global $SU_{\ell + r}(N)$ color-flavor locked symmetry),  part of the $N\, n$  zeromodes is the vortex deformation-translational modes, which connect the $n$ coaxial vortices all in the same orientation to  $n$ far-separated singly wound vortices. The onset of such a transition was shown to be characterized by the formation of the vacuum bubble inside the vortex bag, see Fig.~\ref{holes}.   Such a hole vortex  solution - a sort of mixed phase configuration - was explicitly constructed   in \cite{Bolognesi:2014saa}.

\begin{figure}
\begin{center}
\includegraphics[width=5in]{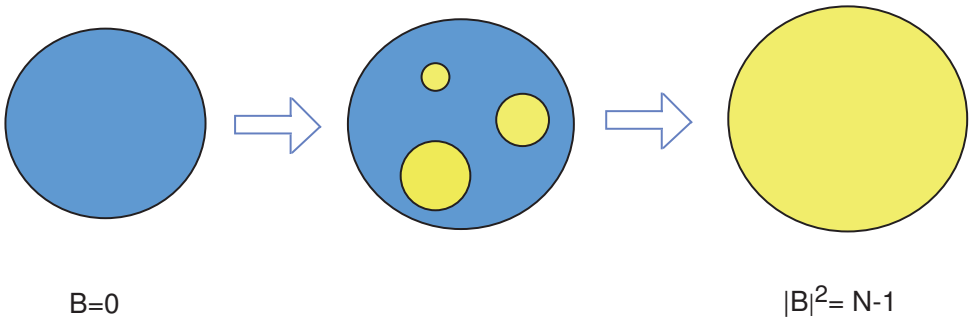}
\caption{The pure $B=0$ vortex (left),  a mixture of  $B=0$ and $|B|^2=N-1$ vortices (center) and the pure $|B|^2= N-1$ vortex (right) are all degenerate. }
\label{Bvortices}
\end{center}
\end{figure}

The transition from   $B=0$ to $|B|^2=N-1$ vortices in the large winding limit may occur in a similar fashion, through the formation of $|B|^2=N-1$ ``bubble vortices"  inside the $B=0$ vortex.   (Fig.~\ref{Bvortices}).
  

\section{Discussion}
\label{cinque}

Generalizing our previous analysis we studied in the first
part of this paper the structure of the nonAbelian  BPS 
vortices in the fully gauged
$U_0(1)\times SU_{\ell}(N) \times SU_r(N)$ theory, with
scalar fields in a bifundamental representation. We
have been able to determine explicitly the vortex
configurations and gauge field mixing, thanks to the fact
that the vortex equations reduce to algebraic
equations in the large winding limit. This allowed us to determine exactly all the vortex zeromodes.
In the case the right group is global, $g_r=0$,
our vortex reduces to the well-understood nonAbelian vortices with
orientational $CP^{N-1}$ moduli. The counting in the large winding limit shows that the dimension of the vortex moduli space in the fully gauged 
theory remains the same, 
$n N$, as in the $g_r=0$ theory. 

Considering the limit of $n$ far separated minimally wound vortices, we conclude that the moduli space dimension for each of them is 
$N= N-1 +1$: consistent with a $CP^{N-1}$ orientational moduli plus translation for each.  This is the same as in the systems in which the right $SU_r(N)$  global symmetry. 

Nevertheless physics of the two systems ($g_r=0$ and $g_r\ne0$) appear to be remarkably different.  In the former case the system enjoys a very nontrivial quantum $CP^{N-1}$ dynamics, which carry on the renormalization group flow below the vortex mass scale, $\sqrt{\xi}$, 
even though the nontrivial physics now takes place  in the $2D$ subspace of the vortex world sheet. The $4D$ bulk is in a completely Higgsed phase and has become sterile, below the vortex mass scale. 

At the moment  the right gauge coupling $g_r$ is turned on,  there appear nontrivial $SU(N)_{\ell+r}$ unbroken $4D$ massless gauge fields,  coupled to the $2D$
massless orientational modes.  Physics changes instantly and qualitatively.  To illustrate some of these new features, we discussed in the second part of this work how some nonlocal, topological effects  make appearance in our systems. Note that these new nonlocal effects are  washed away  in the former case ($g_r=0$),   due to the fact that the $2D$
orientational modes strongly fluctuate at long distances.  In other words, the interesting $2D$ quantum dynamics kills the would-be  beautiful nonlocal effects in $4D$ of the $g_r\ne 0$ theory.  
In this sense,  the two formally  very closely related systems  --  the $U_0(1)\times SU_{\ell}(N) \times SU_r(N)$ theories with $g_r=0$ and $g_r\ne 0$ -- are actually more complementary than similar.  

We have seen that the general theory developed earlier  to deal with vortices with a residual gauge invariance in the bulk, basically applies also to our model. At the same time,  our model presents some interesting features that are new with respect to the specific examples which were treated  in the existing literature. Our  nonAbelian vortex system is 
rather special in that  the unbroken group $H$ is both continuous and nonAbelian.  Each vortex is labeled by an element $\Gamma$ which further breaks $H \to \widetilde{H}$.  The vortex moduli space contains a $CP^{N-1}$ orientational moduli.
We have shown that the effect of such vortex orientational modes  is physical, in spite of the fact that they 
arise from the breaking of the $SU(N)_{\ell+r}$ {\it gauge} symmetry by the vortices,     and visible at large distances  as global topological effects such as the AB scattering. 
This is a particular type of nonAbelian AB effect. As discussed in some  cases earlier,  phenomena such as  vortex-vortex scattering, topological obstruction, Cheshire charge,  and nonAbelian statistics  are  all  present. 

More dynamical aspects of zeromode excitations of  our systems and the properties of the low-energy effective actions,   will be discussed in a separate work.

\section*{Acknowledgments} We thank Roberto Auzzi, Jarah Evslin, Simone Giacomelli, Muneto Nitta and  Keisuke Ohashi for discussions.  The  work of SB is funded by the grant ``Rientro dei Cervelli Rita Levi Montalcini'' of the Italian government.  This work is supported by the INFN special research project, ``Gauge and String Theories'' (GAST).

\end{document}